\documentclass[aps,prb,superscriptaddress,twocolumn,showpacs,amsmath,floatfix,citeautoscript]{revtex4}
\usepackage{graphicx}
\usepackage{float}
\usepackage{dcolumn}
\usepackage{xcolor}
\usepackage{latexsym,bm}
\usepackage[normalem]{ulem}
\usepackage{multirow}
\usepackage{appendix}
\usepackage{amsmath}
\usepackage{amsfonts}
\usepackage{booktabs}
\usepackage{subfigure}
\usepackage{threeparttable}
\usepackage{CJKutf8}
\usepackage{csquotes}

\begin{document}

\hyphenpenalty=5000
\tolerance=1000

\begin{CJK}{UTF8}{gkai}

\title{Multihyperuniformity in high entropy MXenes}

\author{Yu Liu}
\affiliation{HEDPS, CAPT, College of Engineering and School of Physics, Peking University, Beijing 100871, P. R. China}
\affiliation{AI for Science Institute, Beijing 100080, P. R. China}



\author{Mohan Chen}
\email[correspondence sent to: ]{mohanchen@pku.edu.cn}
\affiliation{HEDPS, CAPT, College of Engineering and School of Physics, Peking University, Beijing 100871, P. R. China}
\affiliation{AI for Science Institute, Beijing 100080, P. R. China}

\date{\today}

\begin{abstract}
{MXenes are a large family of two-dimensional transition metal carbides and nitrides that possess excellent electrical conductivity, high volumetric capacitance, great mechanical properties, and hydrophilicity. In this work, we generalize the concept of multihyperuniformity (MH), an exotic state that can exist in a disordered multi-component system, to two-dimensional materials MXenes. Disordered hyperuniform systems possess an isotropic local structure that lacks traditional translational and orientational order, yet they completely suppress infinite-wavelength density fluctuations as in perfect crystals and, in this sense, possess a hidden long-range order. In particular, we evaluate the static structure factor of the individual components present in the high entropy (HE) MXene experimental sample TiVCMoCr based on high-solution SEM imaging data, which suggests this HE MXene system is at least effectively multihyperuniform. We then devise a packing algorithm to generate multihyperuniform models of HE MXene systems. The MH HE MXenes are predicted to be energetically more stable compared to the prevailing (quasi)random models of the HE MXenes due to the hidden long-range order. Moreover, the MH structure exhibits a distinctly smaller lattice distortion, which has a vital effect on the electronic properties of HE MXenes, such as the density of states and charge distribution. This systematic study of HE MXenes strengthens our fundamental understanding of these systems, and suggests possible exotic physical properties, as endowed by the multihyperuniformity.}
\end{abstract}


\maketitle

\end{CJK}

\section{Introduction}

MXenes, a large family of two-dimensional (2D) metal carbides and nitrides with a general formula
of $\mathrm{M_{n+1}X_nT_x~(n = 1\text{-}4)}$, are structured with two or more layers of transition metal (M) atoms arranged in a honeycomb-like 2D lattice, interspersed with carbon and/or nitrogen (X) layers occupying the octahedral sites between adjacent transition metal layers.~\cite{11AM-Naguib,12ACSN-Naguib}
Terminal groups of MXenes such as -OH, -O, and -F, commonly denoted as $\mathrm{T_x}$, are tunable during etching A-layer from the layered hexagonal MAX phases,~\cite{11ARMR-Barsoum,19TC-Sokol} where A is an element mainly from groups 13–16 of the periodic table.
Owing to the unique 2D structure, MXenes exhibit excellent electrical conductivity,~\cite{19ACSAMI-Mirkhani} high volumetric capacitance,~\cite{17NE-Lukatskaya} great mechanical properties,~\cite{18SA-Lipatov} and hydrophilicity.~\cite{20S-Kamysbayev}
Thanks to their outstanding properties, MXenes have been extensively investigated for potential applications in various fields, such as batteries,~\cite{18ACIE-Chen,19NC-Zhang,21AM-Li,18Nanoscale-Meng,18JMCA-Meng,19-Nanoscale-Li,21ACSN-Li} electromagnetic interference shielding,~\cite{16S-Shahzad,20ACSN-Han} electrocatalysis,~\cite{21AFM-Lv,23ACSN-Li,20ACSN-Guo} supercapacitors,~\cite{18N-Xia,22NE-Ma} and so on.

High-entropy materials (HEMs) are a class of materials composed of mixtures of equal or relatively large proportions of multiple principal elements.~\cite{04AEM-Yeh,04MSEA-Cantor}
The presence of several principal elements can decrease Gibbs free energy, benefiting from the entropy stabilization effect, and thus promote the formation of solid-solution phases.
Highly diversified structures enable HEMs to exhibit unexpected mechanical, physical, and chemical properties, offering great promise for applications in mechanics, energy storage, and conversion.~\cite{19AM-Sarkar,20AM-George,20AM-Wang,21AEM-Cui}
Recently, the high entropy (HE) strategy was extended into the MXene design space, further enhancing its tunability.~\cite{21ACSN-Nemani}
HE MXenes have great potential for energy storage and conversion due to their vast configurational space.
For example, Du et al. prepared the HE MXene $\mathrm{(Ti_{1/5}V_{1/5}Zr_{1/5}Nb_{1/5}Ta_{1/5})_2C_xN_{1-x}}$ with high mechanical strains, which shows good adsorption and catalytic activities for lithium polysulfides.~\cite{22AEM-Du}
In 2023, a V-doped Co2P coupled with HE MXene heterostructure catalyst was prepared by a two-step electrodeposition and controlled phosphorization process, showing excellent hydrogen evolution reaction and oxygen evolution reaction activity and long-term stability under alkaline conditions.~\cite{23JEC-Ma}
Recently, Tan et al. proposed a thermodynamic competition strategy and enabled the successful synthesis of a stable HE MXene $\mathrm{Ti_{1.1}V_{1.1}Cr_{0.4}Nb_{1.4}C_3T_x}$ with weight capacitances of 292.74 $\mathrm{F/g}$ at 2 $\mathrm{mV/s}$ and 137.20 $\mathrm{F/g}$ at $\mathrm{mV/s}$.~\cite{24AM-Tan}

Due to the difficulties in preparation, only a few HE MXenes have been synthesized successfully.
The detailed atomic configurations of HE MXenes remain elusive, considering the relatively recent proposition of combining the HE concept with MXenes.
In terms of theoretical calculations of HE MXenes,~\cite{23JMCA-Seong} the dominant models employed are random mixture and special quasirandom structure (SQS) models.~\cite{90L-Zunger,13Calphad-Walle}
These models essentially posit a straightforward assumption that atoms of diverse types within MXene structures are randomly distributed on the sites of an underlying crystalline lattice.
However, we notice that the clustering of atoms of the same type is suppressed as displayed in elemental mapping images of recent experimental studies.~\cite{23JEC-Ma,23AM-He,23SM-Ma}
Given that atoms prefer specific neighboring configurations, which are not captured by random mixture and SQS models, we anticipate that when an observation window is randomly shifted throughout the system, the count of atoms of a particular element within that window would exhibit little fluctuations, especially for larger window sizes.

Hyperuniformity~\cite{03E-Torquato} is a special long-range order characterized by an ordered metric based on particle number variance $ \sigma_N^2(R)$ in a spherical observation window, which vanishes when normalized by the observation window volume (scales as $\sim R^d$ in $d$-dimensional Euclidean space) in the infinite-window limit, i.e.,
\begin{equation}
\lim_{r\rightarrow \infty} \sigma_N^2(R)/R^d = 0.
\end{equation}
Hyperuniformity is equivalently manifested as the
vanishing static structure factor in the infinite-wavelength (or zero-wavenumber) limit, i.e., $\lim_{k\rightarrow 0}S(k) = 0$, where $k$ is the wavenumber. The small-$k$ scaling behavior of $S(k) \sim k^\alpha$ determines the large-$R$
asymptotic behavior of $\sigma_N^2(R)$, based on which all hyperuniform
systems, disordered or not, can be categorized into three classes:
$\sigma_N^2(R) \sim R^{d-1}$ for $\alpha>1$ (class I); $\sigma_N^2(R)
\sim R^{d-1}\ln(R)$ for $\alpha=1$ (class II); and $\sigma_N^2(R)
\sim R^{d-\alpha}$ for $0<\alpha<1$ (class III).~\cite{18PR-Torquato}
Consequently, hyperuniform systems encompass all crystals and quasicrystals.~\cite{03E-Torquato, 18PR-Torquato} 
Crystals belong to a special subset of class-I hyperuniform systems and possess a zero structure factor for a range of wavenumbers starting from the origin, i.e., $S(k) = 0$ for $k<K^*$ (excluding the forward scattering), which are referred to as stealthy hyperuniform systems. 
However, conventional disordered systems like liquids and glasses do not exhibit this characteristic hyperuniformity.
%
%


Disordered hyperuniform (DHU) systems~\cite{03E-Torquato,18PR-Torquato} are recently discovered novel states of many-body systems between a perfect crystal and a liquid. DHU systems, akin to liquids and glasses, exhibit statistical isotropy and manifest an absence of Bragg peaks, which signifies the lack of conventional long-range translational and orientational order.
However, they completely suppress normalized large-scale density fluctuations similar to crystals, thus possessing a hidden long-range order.~\cite{03E-Torquato,18PR-Torquato,09JSM-Zachary} DHU states have been observed across a broad range of equilibrium and nonequilibrium physical and biological systems and seem to confer unique desirable properties (such as complete isotropic photonic and phononic band gaps) that are unattainable in either totally disordered or perfectly crystalline states.~\cite{03E-Torquato,18PR-Torquato,09PNAS-Florescu,13PNAS-Man,14E-Jiao,15L-Hexner,15L-Jack,17E-Xu,18AM-Chen,18L-Chremos,19NC-Klatt,19PNAS-Lei,19SA-Lei,20SA-Zheng,21PNAS-Huang,21PNAS-Chen,21L-Klatt,21B-Zheng,22PA-Jiao,22B-Chen,22B-Chen2, chen2023disordered, liu2024universal, zhuang2024vibrational, mkhonta2024uncovering, bai2024nitrogen}


Multihyperuniformity is an exotic hyperuniform state that can exist in a multi-component system, in which each component is individually hyperuniform, and consequently, the entire system is overall hyperuniform. 
To date, two examples of multihyperuniform (MH) systems have been reported, including the distribution of photoreceptor cones in avian retina~\cite{14E-Jiao} and high-entropy alloys (HEA)~\cite{chen2023multihyperuniform}. 
The emergence of multihyperuniformity in such systems is typically related to certain effective mutual \enquote{exclusion effects} among the particles of the same species. For example, in the case of avian photoreceptors, once a cell's phenotype is determined, an inhibition factor is released to suppress the formation of neighboring cells of the same type.~\cite{14E-Jiao} 
Similarly, in the case of HEAs, enthalpy minimization favors hetero-type neighbor arrangement and suppresses clustering of the same species.~\cite{chen2023disordered} 
Due to the similarity of HEAs and HE MXenes, it is natural to ask the question: \textit{Can HE MXenes also be MH}?    

In this work, we present a comprehensive study of multihyperuniformity in HE MXenes. We first carry out a detailed analysis of high-resolution SEM images via evaluating the static structure factor of the individual components present in the high entropy (HE) MXene experimental sample TiVCMoCr~\cite{23AM-He}, which suggest this HE MXene system is at least effectively multihyperuniform (i.e., that the individual transition metal atomic species tend to distribute hyperuniformly). Motivated by this observation, we generalize a multi-scale packing algorithm~\cite{14E-Jiao} to generate multihyperuniform structural models for HE MXenes.
Using the HE MXene $\mathrm{(TiVZrMo)_2C}$ as an example, we predict that the MH model results in a more energetically stable state compared to the commonly used random mixture configurations. The hidden long-range order in the MH models also leads to a smaller deviation from Vegard’s law~\cite{21ZP-Vegard} and lower lattice distortion, which affects the electronic properties of HE MXenes.

The rest of the paper is organized as follows: 
In Sec.~\ref{method}, we discuss the generalization of the multihyperuniformity concept to the 2D material HE MXenes, and the details of the first-principles calculations employed in this work.
In Sec.~\ref{results1}, we analyze and discuss the multihyperuniformity in the experimental samples.
In the rest of Sec.~\ref{results}, we employ first-principles calculations on $\mathrm{(TiVZrMo)_2C}$, which is used as an example to investigate the structural and electronic properties of HE MXenes with multihyperuniformity.
In Sec.~\ref{Conclusions}, We provide concluding remarks and propose future work based on the integration with machine learning.

\section{Methods}\label{method}

\subsection{Generation of MH patterns on lattices as a multi-scale packing problem} \label{method1}

We generalize the multi-scale packing algorithm~\cite{14E-Jiao} to generate structural models of MH HE MXene samples, whose key statistics match experimental data very well (see Results for details). In particular, we randomly distribute a total of $N = \sum n_s$ atoms of $m$ different types on a prescribed lattice $\Lambda$ in $\mathbb{R}^2$, where $n_s$ is the number of atoms of types $s$, and $s = 1, \ldots, m$. Each lattice site can only host one atom (of any type) and the atoms of the same type repel each other via a soft, finite-ranged interaction, i.e., $\phi_s(r_{ij}) = c_s (r_{ij} - 2R_s)^2$, for $r_{ij} < 2R_s$, and $\phi_s(r_{ij}) = 0$ otherwise, where $r_{ij}$ is the distance between the atoms $i$ and $j$ of type $s$, $R_s$ is a type-dependent range of the soft repulsive interaction, and $c_s$ is an algorithmic parameter which is chosen to be unity for all types. Simulated annealing is then employed to minimize the total energy of the system $E = \sum_s\sum_{i, j} \phi_s(r_{ij})$.~\cite{14E-Jiao} 
This is achieved by randomly switching the positions of a pair of atoms of different types and accepting the new configuration with a probability given by the Metropolis rule, which on average leads to configurations with lower energy and in the meantime avoids the system stuck in shallow minima. The simulation is determined once a global (or very stable local) energy minimum is achieved, and the resulting MH 2D configuration is characterized (see Sec.~\ref{results}) and mapped to a model MXene structure (see Fig.~\ref{fig:structures}). 

In particular, the multi-scale packing algorithm is utilized to generate a single atom layer in multihyperuniformity.
We first generate two MH transition metal atom layers independently and position one layer directly above the other at a specified distance. Next, we shift the upper layer so that its atoms are aligned with the centers of the triangles formed by the atoms in the lower layer. Finally, we insert a layer of carbon atoms, arranged in the same triangular lattice, between the two transition metal layers to achieve the final 2D configuration.
For comparison, we also generate random configurations of the atoms and the associated random mixing model of HE MXenes, which has been widely used in previous numerical studies~\cite{21AM-Du,23DT-Li,23JMCA-Seong}.



\subsection{First-principles calculations}

The Atomic-orbital Based Ab-initio Computation at UStc (ABACUS) v3.5.1 package \cite{10JPCM-Chen,16CMS-Li} are employed to perform all the density functional theory (DFT)~\cite{64-Hohenberg,65-Kohn} calculations.
The norm-conserving pseudopotential~\cite{13B-Hamann,15CPC-Schlipf} is employed to describe the ion-electron interactions.
The valence electron configurations of Ti, V, Zr, Mo, and C are $3s^{2}3p^{6}4s^{2}3d^{2}$, $3s^{2}3p^{6}4s^{2}3d^{3}$, $4s^{2}4p^{6}5s^{2}4d^{2}$, $4s^{2}4p^{6}5s^{2}4d^{4}$, and $2s^{2}2p^{2}$. 
The generalized gradient approximation (GGA) in the form of the Perdew-Burke-Ernzerhof
(PBE) \cite{96L-Perdew} is used for the exchange-correlation functional.
In order to deal with large systems of the HE MXenes $\mathrm{(TiVZrMo)_2C}$ containing 1536 atoms and 15104 valence electrons, we employ numerical atomic orbitals (NAO) in the form of double-$\zeta$ plus polarization function (DZP) orbitals as basis sets in our calculations,
whose accuracy and consistency have been verified in previous studies~\cite{16CMS-Li,21JNM-Liu}.
It has been demonstrated that ABACUS is capable of performing DFT calculations for large systems using NAOs.~\cite{21JNM-Liu,22PCCP-Liu,22B-Chen,24AM-Liu}
Specifically, $4s2p2d1f$ NAOs are utilized for the four transition metal elements, whose radius cutoff is set to 9 Bohr.
Meanwhile, we employ $2s2p1d$ NAOs for C, whose radius cutoff is set to 8 Bohr.
The energy cutoff is set to 60 Ry.
Besides, periodic boundary conditions and a single $k$-point ($\Gamma$ point) are used.
The values of the energy cutoff and $k$-points density ensure energy convergence to within 1 meV/atom.
We employ the Gaussian smearing method with a smearing width of 0.01 Ry.
Structural optimizations are performed with the conjugated gradient method
until all components of the stress tensor are below 2 kbar and forces on each atom are below 0.04 eV/\AA, ensuring the convergence of the total energy of the supercells.

\section{Results and Discussions} \label{results}

\begin{figure*}[htbp]
    \centering
    \includegraphics[width=16cm]{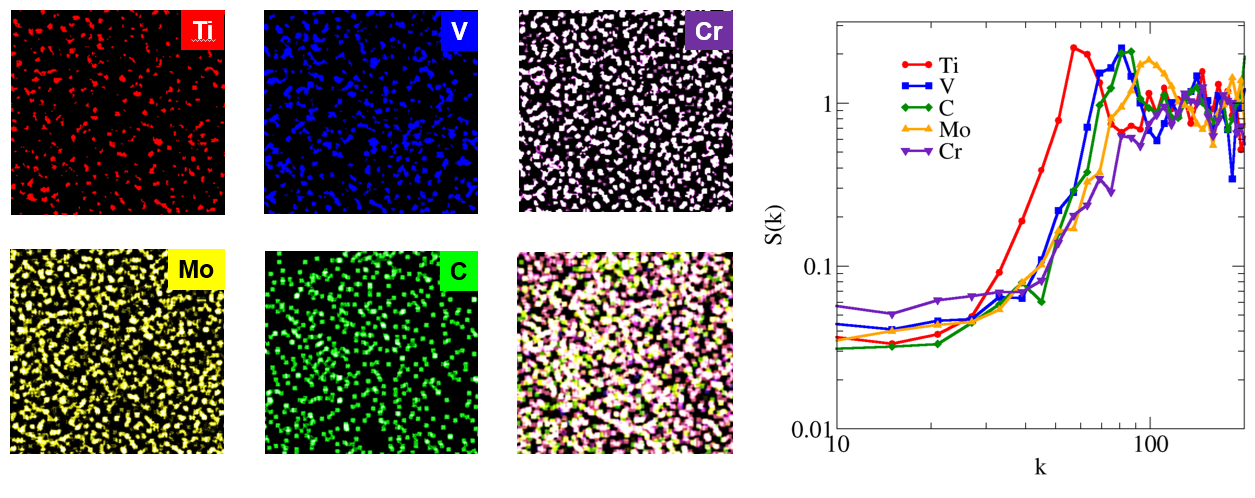}
    \caption{
        (Color online) SEM images~\cite{23AM-He} of $\mathrm{TiVCMoCr}$ HE MXenes sample showing each of the atomic species and the overall distribution (left panels) and the associated static structure factor $S(k)$. The linear size of the box is $L= 20$ nm. The unit of the wavenumber is $0.1~nm^{-1}$.
    }
    \label{fig:exp}
\end{figure*}

\subsection{Multihyperuniformity of HE MXene samples} \label{results1}


Compared to the extensive research on HE 3D crystalline solids, the study of monolayer HE MXenes is still in its early stages, with only a few studies reported so far.~\cite{15NC-Zou}
Given the limited number of reports on HE 2D materials, obtaining a suitable sample with high-resolution images is challenging.
In this work, we analyze the SEM images of the as-prepared $\mathrm{TiVCrMoC}_3$ HE MXene~\cite{23AM-He} to detect the hidden hyperuniformity within the system.
As shown in Fig. 1a, atoms of different types (shown as different colored dots) were identified, and their center positions were extracted and converted to a point configuration. For a single point configuration with $N$ particles at positions $\mathbf{r}^N = (\mathbf{r}_1,\dots,\mathbf{r}_N)$ with periodic boundaries within an orthogonal fundamental cell, the static structure factor ${S}(\mathbf{k})$ is given by
\begin{equation}\label{eq:Skcomp}
    {S}(\mathbf{k}) = \frac{|\sum_{j=1}^N \textrm{exp}(-i\mathbf{k}\cdot\mathbf{r}_j)|^2}{N},
\end{equation}
which is used to compute $S(\mathbf{k})$ directly from the point configurations associated with the atom centers derived from both experiments and simulations.
Figure 1b shows $S(k)$ for all different types of atoms, which indicates a weak power-law behavior $S(k) \sim k^\alpha$ as $k \rightarrow 0$, where the hyperuniformity exponent $\alpha \in (0, 0.08)$ for different atomic species. An extrapolation analysis indicates the hyperuniformity index $H = S(k\rightarrow0)/S_{peak} \approx 10^{-2}$, suggesting the HE MXenes system is at least effectively hyperuniform. We note that the SEM images may contain a curved portion of the samples, which might have degraded the degree of hyperuniformity based on the image analysis, and the original sample might possess an even higher degree of hyperuniformity.

To the best of our knowledge, the as-prepared $\mathrm{TiVCrMoC}_3$ HE MXene~\cite{23AM-He} is the most suitable for statistical analysis within the accessible experimental data.
However, we note that this sample consists of four transition metal layers and three carbon layers, as well as terminal groups.
Additionally, the presence of the antiferromagnetic atom Cr adds complexity to the simulation.
For simplicity, we construct a three-layer HE MXene, $\mathrm{(TiVZrMo)_2C}$, without terminal groups as the simulated system to investigate multihyperuniformity in HE MXenes. Figure 2 shows $S(k)$ of the simulated MH systems with equal molar fractions for all four transition metal species, possessing $\alpha \approx 1.02$ and $S(k=0) \approx 10^{-6}$. 
We note that since the four transition metal species are equal molar and, thus, equivalent in the configurations, we expect their $S(k)$'s are also statistically equivalent. It can be seen that the simulated systems clearly exhibit a higher degree of hyperuniformity compared to the raw image analysis. 
Possible reasons for this discrepancy include limited resolution of the SEM images, curvatures within the material sample as well as presence of defects in the sample. In the subsequent calculations, we will employ the MH structural models generated using the multi-scale packing algorithm described in Sec.~\ref{method1}.


\subsection{Energetic stability and lattice distortions}

\begin{figure*}[htbp]
    \centering
    \includegraphics[width=16cm]{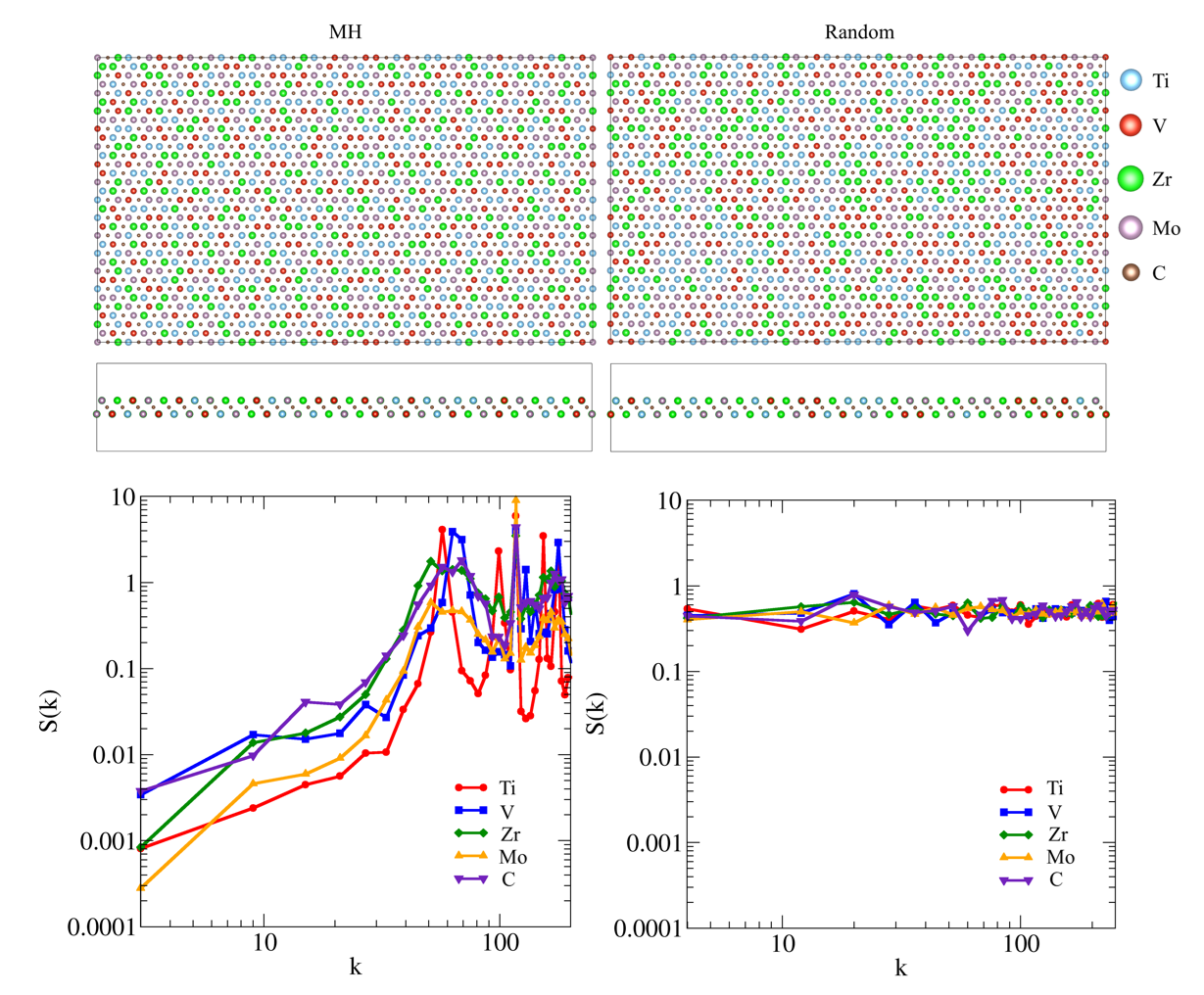}
    \caption{
        (Color online) Top views (top panel), side views (middle panel) and static structure factor $S(k)$ of representative MH (left) and random structures (right) of $\mathrm{(TiVZrMo)_2C}$ HE MXenes containing 1536 atoms in total. The unit of the wavenumber is $0.1~nm^{-1}$
    }
    \label{fig:structures}
\end{figure*}

In order to further validate the energetic stability of the MH HE MXenes, we compute the total energy of the MH and random structures via DFT calculations.
As depicted in Fig.~\ref{fig:structures}, the MH and random structures include 1536 atoms, which are composed of two transition atom layers and one carbon atom layer between them.
To avoid any interactions between adjacent monolayers of HE MXenes, we introduce a 12 \AA~thick vacuum layer along the direction perpendicular to the monolayer's surface.
Notably, we do not construct the SQS model and compute its total energy.
The basic idea of SQS is to generate a minimally sized supercell that approximates a random (disordered) solid solution in the sense that its cluster vectors closely resemble those of truly random systems.
This approach's advantage lies in its ability to simulate random systems with lower computational costs, thereby enhancing the computational efficiency of DFT calculations.
However, we must construct large enough supercells to verify the hidden long-range order in MH HE MXenes.
When the system scale reaches a certain magnitude, the atomic configuration tends to assume a statistically random nature, rendering straightforward stochastic structures capable of representing the system's atomic arrangements.
Consequently, we opt not to utilize SQS models in this research, as their outcomes align with those obtained from random structures.~\cite{23AM-Chen}
Additionally, we perform DFT calculations on various independently generated MH and random structures, observing negligible variations among them. Thus, our subsequent analysis primarily focuses on the representative MH and random structures in Fig.~\ref{fig:structures}.
As shown in Table~\ref{table:energy_structure}, the MH structure exhibits a total energy per atom that is 8.1 meV lower than that of its random counterpart, 
indicating a higher degree of energetic stability in the MH HE MXene compared to the random configuration,
which strongly suggests the existence of a hidden long-range order in HE MXenes.

\begin{table}[htbp]
	\centering
	\caption{The total energy per atom $E$, width $d$, and distortion parameter $\Delta h$ of representative MH and random structures of $\mathrm{(TiVZrMo)_2C}$ HE MXenes optimized via DFT calculations by ABACUS. The width $d$ depicts the distance between the center of the two transition metal layers. The distortion parameter $\Delta h$ is defined to be the average width of the transition metal layers.}
	\label{table:energy_structure}
	\setlength{\tabcolsep}{8pt}
	\renewcommand\arraystretch{1.5}
	\begin{tabular}{ccc}
		\toprule
		\hline
		Structures & MH & Random \\
		\midrule
		\hline
        $E$ (eV/atom) & -1161.4439 & -1161.4358 \\
		$d$ (\AA) & 2.319 & 2.310 \\
        $\Delta h$ (\AA) & 0.613 & 1.286 \\
		\bottomrule
		\hline
	\end{tabular}
\end{table}

We predict that MH structures with suppressed composition fluctuations exhibit a smaller deviation from Vegard’s law than random structures. 
As displayed in Table~\ref{table:energy_structure}, the width $d$ (i.e., the distance between the center of the two transition metal layers) of the MH and random structures of $\mathrm{(TiVZrMo)_2C}$ HE MXenes optimized by DFT calculations is 2.319 and 2.310 \AA, respectively.
According to Vegard’s law, the width $d$ of an equimolar $\mathrm{(TiVZrMo)_2C}$ HE MXene is predicted to be 2.327 \AA, the weighted average width $d$ of pure $\mathrm{Ti_2C}$ (2.309 \AA), $\mathrm{V_2C}$ (2.175 \AA), $\mathrm{Zr_2C}$ (2.543 \AA), and $\mathrm{Mo_2C}$ (2.280 \AA) MXenes.
The results indeed validate our assumption that the MH structure exhibits a smaller deviation from Vegard’s law and reflects the nearly ideal mixing of multiple transition metal elements in the HE MXenes.

In single transition metal MXenes, the transition metal atoms within the same layer are positioned on a parallel horizontal plane.
However, owing to the disparities in atomic sizes and the intricate interplay among multiple transition metal elements in HE MXenes, these transition metal atoms exhibit a certain degree of displacement, leading to the loss of planarity in the atomic layers.
Thus, we also evaluate the lattice distortion via a metric $\Delta h$, which is defined to be the average width of the transition metal layers.
We find that the distortion parameter $\Delta h$ of the MH structure (0.613 \AA) is much less than that of the random structure (1.286 \AA).
The MH configuration exhibits a distinctly smaller lattice distortion, resulting in a more energetically stable state compared to its random counterparts. 
The enhanced stability of the MH system can be attributed to the suppressed composition fluctuations.
In terms of the relationship between lattice distortion and energy storage applications,
previous studies suggested that the strong strains induced by lattice distortion enable to effectively guide the nucleation and uniform growth of dendrite-free lithium on HE MXene layers, delivering a stable cycling performance and good deep stripping-plating levels.~\cite{21AM-Du}
Such a vital correlation emphasizes the importance of structural modeling methods in theoretical simulations.
In this respect, the more stable MH structures predict more accurate computational results than the random structures for HE MXenes.

\subsection{Density of states}

\begin{figure}[htbp]
    \centering
    \includegraphics[width=8cm]{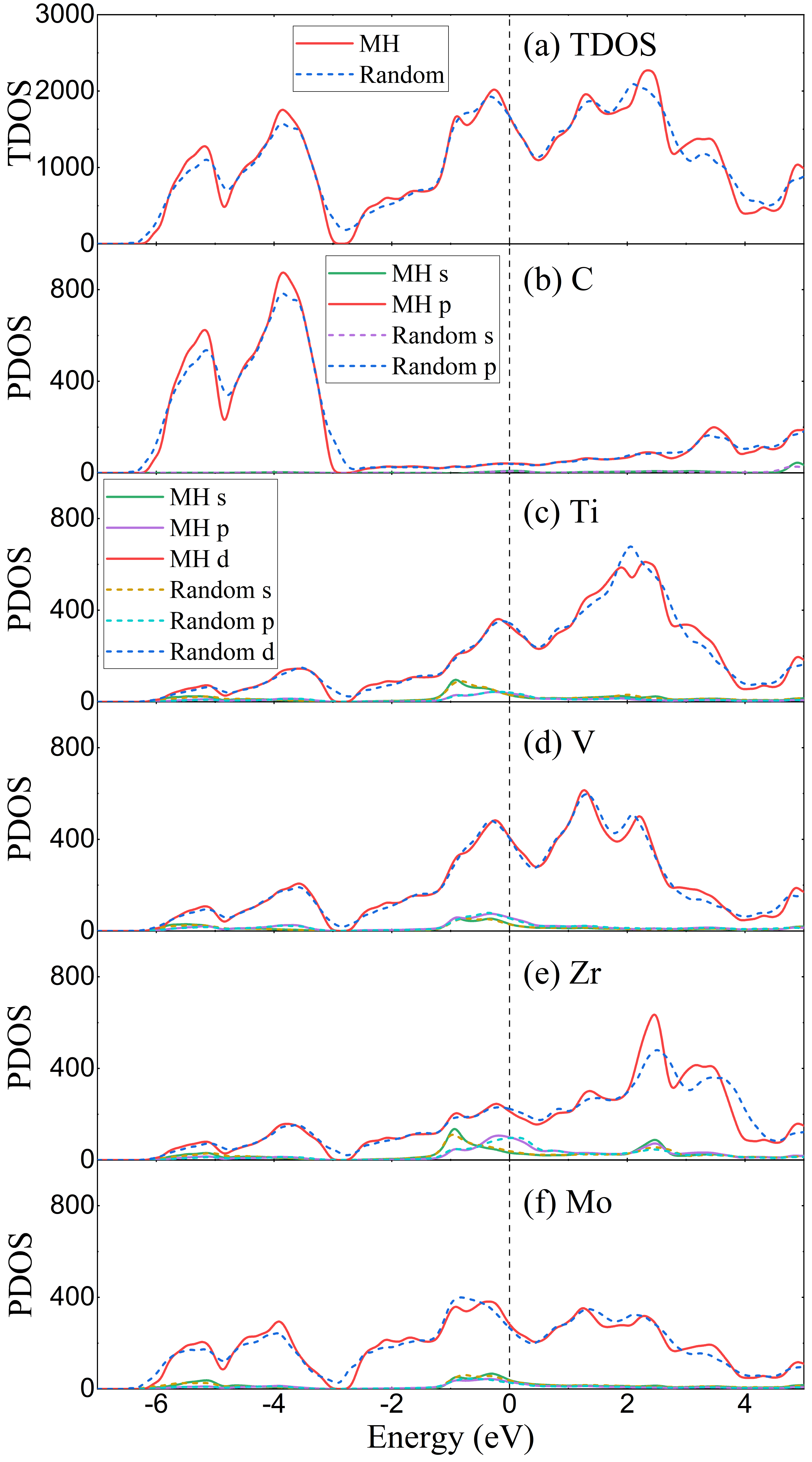}
    \caption{
        (Color online) Density of states (top panel) and projected density of states (bottom panel) of representative MH and random structures of $\mathrm{(TiVZrMo)_2C}$ HE MXenes. The Fermi level is shifted to 0 eV.
    }
    \label{fig:dos}
\end{figure}

\begin{figure}[htbp]
    \centering
    \includegraphics[width=8cm]{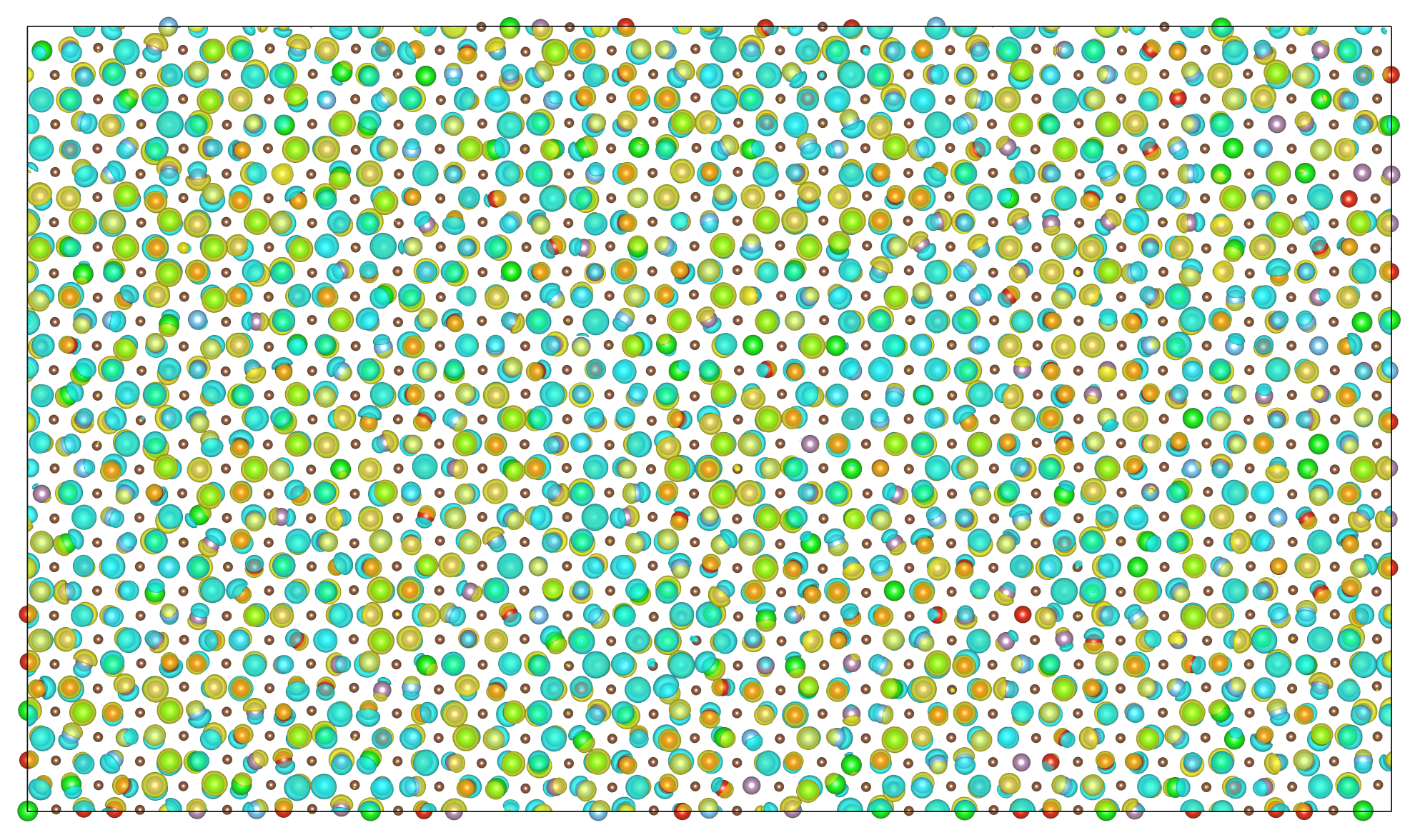}
    \caption{
        (Color online) The charge density difference $\Delta \rho$ between the MH and random structures of $\mathrm{(TiVZrMo)_2C}$ HE MXenes. The isosurface value is set to 0.15.
    }
    \label{fig:charge_density}
\end{figure}

We also investigate the effect of MH long-range order on the electronic structure of the $\mathrm{(TiVZrMo)_2C}$ HE MXenes. 
The density of states (DOS) and projected density of states (PDOS) of representative MH and random structures of $\mathrm{(TiVZrMo)_2C}$ HE MXenes are plotted in Fig.~\ref{fig:dos}.
Notably, we observe that the gap, present at approximately -3 eV below the Fermi level in the DOS of the MH configuration, is absent in that of the random structure.
Prior researches indicate that mechanical strain influences the electronic properties of graphene, leading to the opening and closing of bandgaps,~\cite{08B-Gui,18N-Yankowitz} 
which is similar to our findings in the context of HE MXenes. 
In the random structure for the $\mathrm{(TiVZrMo)_2C}$ HE MXene, large lattice distortion arises from the local atom clustering of the same atomic species,
which is suppressed due to the presence of long-range order inherent to the MH model.
As an intuitive manifestation, we plot the charge density difference $\Delta \rho$ between the MH and random structures of $\mathrm{(TiVZrMo)_2C}$ HE MXenes.
We observe that the distribution of charge density in the random structure is not uniform compared to that of the MH structure due to the local atom clustering of the same atomic species.
According to the PDOS panel in Fig.~\ref{fig:dos}, DOS around the Fermi level arises mainly from the $d$ orbitals of the transition metal atoms, suggesting that the electric conductivity is mainly attributed to the $d$ electrons in the outer region.
In addition, we note that the $d$ orbitals of transition metal atoms and the $p$ orbitals of carbon atoms overlap from -6 to -3 eV below the Fermi level, indicating strong hybridization between them.

\subsection{Bader charge analysis}

Finally, we investigate the charge transfer of multiple atomic species in the MH and random structures of $\mathrm{(TiVZrMo)_2C}$ HE MXenes using the Bader charge analysis method.~\cite{06CMS-Henkelman,07JCC-Sanville,09JPCM-Tang}
As shown in Table~\ref{table:bader_charge}, carbon atoms gain valence electrons, whereas all transition metal atoms lose valence electrons due to the much stronger electronegativity of carbon atoms than transition metals.
The transition metal atoms mainly donate $d$ electrons, which bond with $p$ electrons of C atoms, thus causing the strong hybridization between them, as discussed in the above section.
The average value of the change in the valence electron number $\Delta n$ is in the same order as the electronegativity, i.e., Zr (1.33), Ti (1.54), V (1.63), Mo (2.16), and C (2.55).
We find that the MH long-range order has a considerable effect on the charge transfer of $\mathrm{(TiVZrMo)_2C}$ HE MXenes.
Taking Zr as an example, $ \Delta n$ (1.145) in the MH structure is larger than that (1.100) in the random structure.
The local clustering of Zr in the random structure leads to a local atomic environment composed of Zr and C atoms.
However, in the MH structure, the MH long-range order suppresses local clustering, resulting in the presence of other transition metals in the neighboring sites of Zr.
Consequently, Zr atoms, having the lowest electronegativity, tend to lose extra valence electrons, which are then transferred to neighboring transition metal atoms of other types.
Similarly, charge transfer in other transition metals can be understood in the same way.
In addition, we metric the fluctuation of valence states with the standard deviation $\sigma$ of the number of valence electrons in different atoms of the same element.
Notably, $\sigma$ in the random structure is larger than that in the MH structure.
The suppressed fluctuation of valence states in the MH structure can once again attributed to the MH long-range order compared to the random structure
since the suppressed local clustering of the same atomic species leads to a more stable local atomic environment.

\begin{table}[htbp]
	\centering
	\caption{The Bader charge analysis of representative MH and random structures of $\mathrm{(TiVZrMo)_2C}$ HE MXenes. $\Delta n$ represents the average value of the change in the valence electron number, and $\sigma$ represents the standard deviation of the number of valence electrons in different atoms of the same element.
		}
	\label{table:bader_charge}
	\setlength{\tabcolsep}{6pt}
	\renewcommand\arraystretch{1.5}
	\begin{tabular}{cccccc}
		\toprule
		\hline
		\multirow{2}*{Elements} & \multicolumn{2}{c}{MH} & & \multicolumn{2}{c}{Random} \\
        \cline{2-3}  \cline{5-6}
        & $\Delta n$ & $\sigma$ & & $\Delta n$ & $\sigma$ \\
		\midrule
		\hline
        Ti & 1.097 & 0.020 & & 1.090 & 0.039 \\
        V & 0.777 & 0.022 & & 0.797 & 0.047 \\
        Zr & 1.145 & 0.029 & & 1.100 & 0.062 \\
        Mo & 0.322 & 0.031 & & 0.380 & 0.072 \\
        C & -1.671 & 0.064 & & -1.684 & 0.103 \\
		\bottomrule
		\hline
	\end{tabular}
\end{table}

\section{Conclusions} \label{Conclusions}

In summary, we presented a novel MH model for HE MXenes and exemplified it through the $\mathrm{(TiVZrMo)_2C}$ HE MXene.
Due to the hidden long-range order, the MH model leads to lower energy and smaller deviation from Vegard’s law compared to the prevailing (quasi)random models of the HE MXenes.
The results align with our analysis of the elemental mapping images~\cite{23AM-He} obtained from the high-angle annular dark-field scanning transmission electron microscopy combined with X-ray energy dispersive spectroscopy, showing that transition metal atoms tend to distribute hyperuniformly.
Additionally, the MH structure exhibits a distinctly smaller lattice distortion, which has a vital effect on the electronic properties of HE MXenes such as DOS and charge distribution.
Previous experimental research suggested that the strong strains in the HE MXene atomic layers are beneficial for the nucleation and growth of lithium in lithium-ion battery applications.~\cite{21AM-Du}
These findings imply that our MH model may account for some of the major discrepancies. 
For example, the practical lithium storage capacity of MXenes falls short of its theoretical value.

It is also important to explore the electrochemical properties of HE MXenes and their potential in energy storage applications.
However, the complexity of the surface environment leads to abundant adsorption sites for ions such as lithium and zinc.
Due to the MH long-range order, large-scale systems for HE MXenes cause an extremely high computational cost for DFT calculations.
In future work, we plan to develop machine-learning potentials.
Capitalizing on its low computational overhead, the machine-learning potential enables us to perform an extensive investigation to assess the impact of multihyperuniformity on the lithium storage capacity of MXenes.

\section*{ACKNOWLEDGMENTS}

Y. L and M. C are supported by the National Natural Science Foundation of China under Grant No.12122401, No.12074007, and No.12135002.
The numerical simulations were performed on the High Performance Computing Platform of CAPT.

\bibliographystyle{apsrev}
\bibliography{ref}
\end{document}